\documentclass[aps,prb,twocolumn,preprintnumbers,amsmath,amssymb,nofootinbib,floatfix]{revtex4}
\pdfoutput=1
\usepackage{graphicx}		
\usepackage{dcolumn}		
\usepackage{bm}			

\begin{document}

\title{Theoretical investigation of magnetic order in ReOFeAs, Re = Ce, Pr }
\author{H. M. Alyahyaei}
\author{R. A. Jishi}
\affiliation{ Department of Physics, California State University, Los Angeles, California 90032 }

\date{\today}

\begin{abstract}
Density functional theory (DFT) calculations are carried out on ReOFeAs, Re = Ce, Pr, the parent compounds of 
the high-T$_c$ superconductors ReO$_{1-x}$F$_{x}$FeAs, in order to determine the magnetic order of the 
ground state. It is found that the magnetic moments on the Fe sites adopt a collinear 
antiferromagnetic order, similar to the case of LaOFeAs. Within the generalized gradient 
approximation along with Coulomb onsite repulsion (GGA+U), we show that the Re magnetic moments also 
adopt an antiferromagnetic order for which, within the ReO layer, same spin Re sites lie along a zigzag
line perpendicular to the Fe spin stripes. While within GGA the Re 4f band crosses the Fermi level, upon inclusion of onsite Coulomb interaction
the 4f band splits and moves away from the Fermi level, making ReOFeAs a Mott insulator.

\end{abstract}

\maketitle

\section{\label{sec:introduction}Introduction}

Recently, a new class of layered, iron-based, high temperature superconductors, has been discovered. Kamihara et al.~\cite{Kamihara:2008} reported a 
superconducting transition temperature T$_c$$=26$ K in fluorine doped LaOFeAs. This is a member 
of a family of compounds known as quaternary oxypnictides with a general formula LnOMPn, where Ln is a lanthanide 
(La, Ce, Pr, ...), M is a transition metal (Mn, Fe, Co, ...) and Pn is a pnicogen (P, As, ...). Shortly afterwards, 
it was shown~\cite{Takahashi:2008} that under pressure the transition temperature increased to 43 K. Hole doping, achieved by 
replacing trivalent La with divalent Sr gave a compound with T$_c$$=25$ K.~\cite{Wen:2008} Replacement of La with other rare 
earth elements gave a series of superconducting compounds ReO$_{1-x}$F$_x$FeAs with Re = Ce, Pr, Nd, or Sm, with 
transition temperatures close to or exceeding 50 K.~\cite{Zocco:2008,Ren:2008,Chen:2008,G_Chen:2008,Z_Ren:2008} Using high pressure techniques, fluorine-free but
oxygen deficient samples were synthesized and found to superconduct at 55 K.~\cite{Ren_2:2008}

\begin{figure}
   \includegraphics[width=0.25\textwidth]{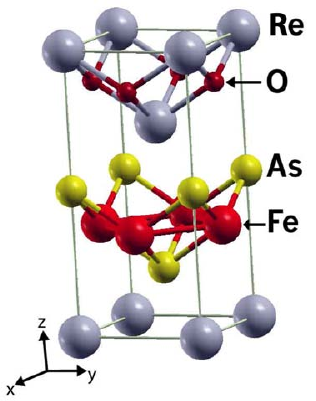}
   \caption{\label{fig:figure1}(Color online) The high-temperature teteragonal unit cell of ReOFeAs, where Re stands for a rare earth atom.}
\end{figure}
The parent compound, ReOFeAs, is a layered compound consisting of a stack of alternating ReO and FeAs layers. At high temperatures, 
the crystal structure is tetragonal with space group P4/nmm and a unit cell, shown in Fig.~\ref{fig:figure1},
that contains two molecules. But at low temperatures, the crystal undergoes a structural phase transition to an 
orthorhombic phase with Cmma space group and a unit cell that contains four molecules.
 The FeAs layer consists of a square planar sheet
of Fe sandwiched between two sheets of As. Upon fluorine doping these compounds become superconductors. It is not 
understood at this stage what mechanism lies behind superconductivity in these compounds. Understanding the electronic
structure of the undoped parent compounds is necessary to understand the doped compounds, especially that in these
iron-based compounds, there is an interplay between magnetism and superconductivity, as is the case in the high-T$_c$ 
cuprates.

Initial calculations using density functional theory (DFT) concluded that LaOFeAs is metallic and nonmagnetic but with
 possible antiferromagnetic (AFM) fluctuations.~\cite{Eschrig:2008,Xu:2008,Haule:2008} More extensive calculations on states with various 
possible magnetic orders in LaOFeAs, however, showed that the magnetic moments of the Fe ions are ordered 
antiferromagnetically in a stripe-like pattern in the Fe plane, resulting in a magnetic unit cell with $\sqrt{2}{a}$x$\sqrt{2}{a}$x$c$ supercell structure,
in contrast to the nuclear axaxc unit cell.~\cite{Ishibashi:2008,Yildirim:2008,Dong:2008,Ma:2008} Indeed, neutron scattering measurements on LaOFeAs reveal the 
existence of such a collinear AFM state at temperatures below 137 K.~\cite{Cruz:2008} 

In this work we study the electronic structure of the parent compounds ReOFeAs, Re = Ce, Pr, using DFT within the generalized gradient 
approximation (GGA). We consider various possible magnetic orders of the Fe and Re ions. We show that in the ground 
state the magnetic moments of the Fe ions adopt a collinear AFM order as in the case of LaOFeAs. Within GGA, the Ce sites 
are paramagnetic, but when the onsite Coulomb interaction is taken into account (GGA+U), the magnetic moments on the 
Ce sites, resulting from the 4f electrons, also adopt an AFM order with a zigzag-like pattern. On the other hand, within
both GGA and GGA+U, the magnetic moments on the Pr sites adopt an AFM order similar to that on the Ce sites.  

\section{\label{sec:method}Method}
The first-principles calculations presented in this work were performed using the all-electron full potential linear augmented plane wave plus local orbitals (FP-LAPW+lo) method as implemented in WIEN2K code~\cite{Blaha_Schwarz:2001}.  In this method the core states are treated in a fully relativistic way but the valence states are treated at a scalar relativistic level.  The exchange-correlation potential was calculated using the generalized gradient approximation (GGA) as proposed by Pedrew, Burke, and Ernzerhof (PBE)~\cite{Perdew_Burke:1996}.

For calculations in this work, the crystal is taken to be orthorhombic, being the low temperature phase, with 
space group C$_{mma}$. The lattice constants are $a = 5.66263$ \AA, $b = 5.63273$ \AA, $c = 8.6444$ \AA,  for the case 
of CeOFeAs~\cite{Zhao:2008}, and $a = 5.6374$ \AA, $b = 5.6063$ \AA, $c = 8.5966$ \AA,  for the case of PrOFeAs~\cite{J. Zhao:2008}.  
 In the low-temperature orthorhombic phase, the unit cell has four Re atoms with crystal
 coordinates Re1 (0, 0.25, z), Re2 (0, 0.75, -z), Re3 (0.5, 0.25, -z), and 
Re4 (0.5, 0.75, z), where z = 0.1402 or 0.1385 for Re = Ce or Pr, respectively. Re1 and Re4 belong to an 
Re-plane above the O-plane, while Re2 and Re3 belong to an Re-plane below the O-plane. These two Re-planes, along with 
the O-plane sandwiched between them, constitute the ReO layer.
 The radii of the muffin-tin spheres are chosen so that the nearby muffin-tin spheres are almost touching. For all structures considered in this work we set the
 parameter R$_{\text{MT}}$K$_{\text{max}}$=7, where R$_{\text{MT}}$ is the smallest muffin-tin radius
 and K$_{\text{max}}$ is a cutoff wave vector.  The valence wave functions inside the muffin-tin spheres are expanded 
in terms of spherical harmonics up to $l_{\text{max}}$ = 10, while in the interstitial region they are expanded 
in plane waves with a wave vector cutoff K$_{\text{max}}$, and the charge density is Fourier expanded up 
to G$_{\text{max}}$=13a$_0^{-1}$, where a$_0$ is the Bohr radius.
  Convergence of the self consistent field calculations is attained with a strict 
charge convergence tolerance of 0.00001 e.
\begin{table*}
        \caption{\label{tab:method_phase_energy}The relative energies per Re ion in five different magnetic orders of the Re ions in ReOFeAs. In all these phases, the paramagnetic (PM), ferromagnetic (FM), antiferromagnetic (AFM), zigzag-along-a antiferromagnetic (z-a-AFM), and zigzag-along-b antiferromagnetic (z-b-AFM), the magnetic order refers only to the magnetic moments on the Re sites. In all these phases, the magnetic moments of the Fe ions are ordered in a collinear antiferromagnetic (c-AFM) fashion with Fe spin stripes in the Fe-plane taken to be parallel to the $b$-axis. The zero of energy corresponds to the case where the Re ion is nonmagnetic. Here U$^{'}$= U-J, where U is the onsite Coulomb interaction and J is the exchange coupling.}
        \begin{ruledtabular}
        \begin{tabular}{l|c c c c c c}
\multicolumn{3}{c}{\ } & \multicolumn{2}{c}{Energy (eV / Re)}&\multicolumn{2}{c}{\ }   \\\hline
 {Phase} &PM    &FM    &AFM   &z-b-AFM &z-a-AFM\\ \hline
   Re = Ce &  &      &      &      &      \\
   GGA &-0.168&0.0056&-0.0127&-0.0054&-0.00623 \\ 
   GGA+U, U$^{'}$= 3 eV &-0.168&-0.489&-0.528&-0.538&-0.583  \\
   GGA+U, U$^{'}$= 5.0 eV &-0.168&-0.452&-0.490&-0.494&-0.551 \\ \hline 
   Re = Pr &  &      &      &      &      \\
   GGA &-0.193&-0.466&-0.461&-0.460&-0.473 \\
   GGA+U, U$^{'}$= 3 eV &-0.193&-0.504&-0.534&-0.523&-0.534 \\
   GGA+U, U$^{'}$= 5.0 eV &-0.193&-0.531&-0.550&-0.532&-0.552        \\

        \end{tabular}
        \end{ruledtabular}
\end{table*}
\section{\label{sec:results_and_discussion}Results and Discussion}
To begin with, we consider within GGA, various magnetic orders of the Fe magnetic moments in the Fe plane: 
nonmagnetic, ferromagnetic (FM), antiferromagnetic (AFM), and collinear antiferromagnetic (c-AFM). We find that the 
c-AFM order of the Fe magnetic moments has the lowest energy.
 The energy of the c-AFM order of the Fe moments is lower that the 
AFM order by $0.031$ $eV$ per Fe atom, lower than the FM order by $0.156$ $eV$ per Fe atom, and lower than the nonmagnetic phase 
by $0.146$ $eV$ per Fe atom. In the AFM order, every spin-up Fe site is surrounded by four nearest neighbor (NN) spin-down Fe 
sites, whereas in the c-AFM order, among the four NN Fe sites surrounding a spin-up Fe site, two along the b-axis are spin-up 
and the other two along the a-axis are spin-down, but the four next nearest neighbor (NNN)
 Fe sites are all spin-down. The AFM and the c-AFM orders 
within the Fe plane were described elsewhere~\cite{Ishibashi:2008,Yildirim:2008,Dong:2008,Ma:2008,Cruz:2008} in connection
with LaOFeAS. That the ground state has a c-AFM order of the Fe magnetic moments, in CeOFeAs and PrOFeAs, is consistent 
with what has already been found in LaOFeAs~\cite{Ishibashi:2008,Yildirim:2008,Dong:2008,Ma:2008,Cruz:2008}, and with 
neutron diffraction measurement on these crystals.~\cite{Zhao:2008,J. Zhao:2008}  

In the remaining calculations we fix the spin order within the Fe-plane to be c-AFM, with the Fe spin stripes taken to be  
along the $b$-direction in the magnetic unit cell. A spin-up stripe in the $a-b$ plane is a line of Fe ions, parallel
to the $b-$axis, with up-spins; this is surrounded in the $a-b$ plane by two spin-down stripes, also parallel to 
the $b-$axis. Fixing the spin order in the Fe-plane,
  we now focus our attention on the 
spin order of the Re ions. Considering only the Re sites in ReOFeAs, we note that every Re site (for example, Ce1)
 has four NN Re sites  
 (2 Ce2 sites at a distance of 3.72 \AA~ and 2 Ce3 sites at a distance of 3.73 \AA)
 and 4 NNN Re sites (4 Ce4 sites at a distance of 3.998 \AA). Thus for
 the magnetic order on the Re sites we 
consider six different cases:

1) The nonmagnetic (NM) phase, where the magnetic moment on every Re site is constrained to be zero.

2) The paramagnetic (PM) phase, where the magnetic moment on each Re site is non zero, but the spins on 
different Re sites are not correlated. 

3) The ferromagnetic (FM) order, where the magnetic moments on all Re sites are aligned in the same direction. 

4) The antiferromagnetic order (AFM), where the four Re ions
in the unit cell, Re1-Re2-Re3-Re4, whose crystal coordinates were given earlier, have the spin arrangement uddu, 
where u stands for up and d stands for down. In
this phase, considering an ReO layer (one O-plane surrounded by two Re-planes) the Re spins in one plane are all up, while
the Re spins in the other plane are all down. That is, in each Re-plane the order is ferromagnetic (FM), but the magnetization
in one Re-plane is opposite to that in the nearby Re-plane lying across from the O-plane.  

\begin{figure}
   \includegraphics[width=0.45\textwidth]{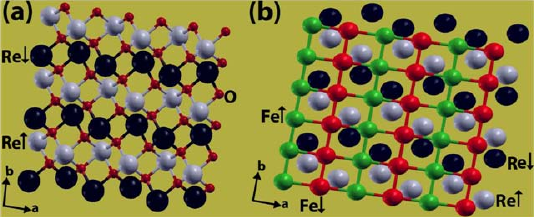}
   \caption{\label{fig:figure2}(Color online) The \{Fe: c-AFM; Re: z-a-AFM\} magnetic phase of ReOFeAs. In (a), a single ReO layer is shown.
The gray balls represent spin-up Re ions while the black ones represent spin-down Re ions. The spin-up Re ions (as well as the spin-down Re ions)
run along a zigzag chain parallel to the $a$-axis. In (b) we show a different perspective of the same magnetic phase. Here we show only
the Fe and Re ions. The Fe spin stripes are along the $b$-direction while the Re spin stripes are "wrinkled" to become zigzag chains along the $a$-direction.}
\end{figure}

5) The zigzag-along-a antiferromagnetic (z-a-AFM) order where a spin-up Re site has two NN spin-up Re sites, two NN spin-down 
Re sites, and four NNN spin-down Re sites. The four Re ions in the unit cell, Re1-Re2-Re3-Re4, have the spin arrangement udud.
Here, if we connect the same-spin NN Re ions in a given ReO layer, we obtain a zigzag chain running parallel 
to the $a$-axis direction, as shown in Fig.~\ref{fig:figure2}.
In this phase, if we consider a single Re-plane (for example, the one containing Re1 and Re4), then it is clear that the Re magnetic
moments in this plane adopt a simple AFM order where each spin-up Re ion (for example Re1) is surrounded by 4 spin-down Re ions (4 Re4 ions).
So in each of the two Re-planes surrounding the O-plane in the ReO layer, the magnetic order is AFM in such a way as to produce same-spin
zigzag chains running along the $a$-direction.

6) The zigzag-along-b antiferromagnetic (z-b-AFM) order for which the four Re ions in the unit magnetic cell,
 Re1-Re2-Re3-Re4, have the spin arrangement uudd.
This is similar to case 5 except that the same-spin Re ions lie on zigzag chains running along the $b$-direction.
 In an isolated ReO layer, and assuming that Re1-Re2 separation is the same as Re1-Re3 separation, this phase
will have the same energy as the previous z-a-AFM phase. But because in the orthorhombic 
structure Re1-Re2 separation is slightly less
than Re1-Re3 separation, and the Fe-plane has spin stripes along the $b$-axis, it follows that these two phases will not
be degenerate, particularly if the Fe and Re spins interact.
  
We make the following two remarks about how the calculations are performed :

i) It is not really possible to calculate directly the total energy of the paramagnetic (PM) phase, because in this 
phase the magnetic moments on the Re sites are randomly oriented; it follows that a unit cell in the (PM) phase will 
contain a very large number of Re atoms. On the other hand, the average magnetic moment per Re site is zero in the 
PM phase. Therefore, one way to calculate the total energy of the PM phase is to constrain the magnetic moment on every Re 
site to be zero; this will make the total energy of the PM phase coincide with that of the nonmagnetic (NM) phase, 
which is not the case in reality. To get around this problem, we note that in the high temperature $\gamma-$phase 
of elemental Cerium crystal, the NN distance between Ce atoms is 3.65 \AA~\cite{Koskenmaki:1981} and that in this phase, while the 6s and 5d 
valence electrons are itinerant, the 4f electron is localized on the Ce site, giving rise to a localized magnetic moment.  
 In CeOFeAs, The distance between the NN Ce sites is 3.72~\AA, slightly larger than in $\gamma-$phase Ce; hence we expect that here the 4f electrons are strongly localized on the Ce sites, giving rise to 
magnetic moments localized on these sites. We calculate the total energy of an isolated Ce atom for both cases 
when the atom is nonmagnetic (half the 4f$^{1}$ electron is spin-up, the other half is spin-down) and when it is 
magnetic. We find that the energy difference is   
\[ 
                E_{magnetic}(Ce)-E_{nonmagnetic}(Ce)= -0.168~ eV.
\]
Therefore, we make the reasonable assumption that 0.168 eV/Ce ion approximates the energy difference between the 
energies of CeOFeAs in the PM phase (Fe ions have c-AFM order but Ce ions are paramagnetic) and in the NM phase 
(Fe ions have c-AFM order while Ce ions are nonmagnetic). Similarly, the energy of the magnetic Pr atom is found to be lower than 
that of the nonmagnetic one by 0.193 eV. 

\begin{figure}
   \includegraphics[width=0.45\textwidth]{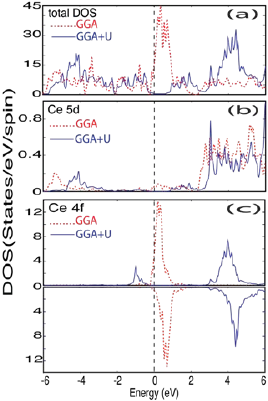}
   \caption{\label{fig:figure3}(Color online) spin-resolved density of states (DOS) in CeOFeAs in
the \{Fe: c-AFM; Ce: z-a-AFM\} phase, within GGA (red), and GGA+U (blue). The zero energy is the Fermi energy.
 In (a) the total DOS is shown, while (b) and (c) show the orbital-resolved atomic DOS
 due to Ce 5d and 4f states, respectively. In (c) the upper panel displays the spin-up DOS,
 while the lower one displays the spin-down DOS. Note that within GGA+U, a
gap opens up at the Fermi energy.}
\end{figure}
\begin{figure}
   \includegraphics[width=0.45\textwidth]{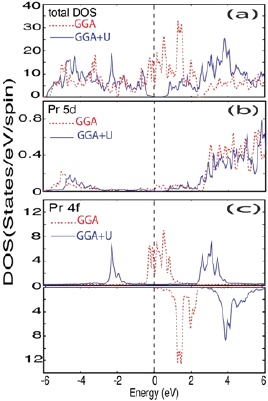}
   \caption{\label{fig:figure4}(Color online)spin-resolved density of states (DOS) in
 PrOFeAs in the \{Fe: c-AFM; Pr: z-a-AFM\} phase,
within GGA (red) and GGA+U (blue). the zero energy is the Fermi energy.
 In (a) the total DOS
is shown, while (b) and (c) show the orbital-resolved atomic DOS due to the Pr 5d and 4f states, respectively.
In (c) the upper panel displays the spin-up DOS, while the lower one displays the spin-down DOS.
 Within GGA+U, a gap opens up
at the Fermi level.}
\end{figure}

ii) In doing the GGA+U calculations, we need the value of U-J, where U is the onsite Coulomb repulsion and J is the 
exchange coupling. For Fe, J$\simeq0.9$ eV, and U has an empirical value in the range 3.5-5.1 eV~\cite{Anisimov:1991}; in our calculation we take U-J$=3.4$ eV 
for the Fe ions. For the Ce ions, values for U-J ranging from $2-5$ eV for GGA calculations are found in the 
literature,~\cite{Cococcioni:2005, Anderson:2007, Loschen:2007} though the value U-J$=5$ eV appears to give better results in describing the electronic structure 
of cerium oxides. For Pr, values as large as 6 eV, we reported for U-J in some oxide of Pr.~\cite{Tran:2008} In the 
calculations reported here we consider two cases where U-J is taken to be 3 eV or 5 eV for both Ce and Pr.

A summary of the total energy calculations is 
given in Table~\ref{tab:method_phase_energy}, where we report the differences in the total energy among the six cases listed earlier. We take the energy of 
the NM phase, in which the Fe moments adopt c-AFM order and the Ce ions are nonmagnetic, as our zero energy. 
The results in Table~\ref{tab:method_phase_energy} show that within GGA, in the absence of the onsite Coulomb interaction, the ground state of CeOFeAs is one 
where the Fe magnetic moments adopt c-AFM order while the Ce sites are paramagnetic. In the presence of onsite Coulomb interaction, 
on the other hand, the Fe magnetic moments adopt the c-AFM order and the Ce magnetic moments the z-a-AFM order.
For the case of PrOFeAs, our results indicate that the Pr magnetic moments also adopt the z-a-AFM order both
within GGA and GGA+U. The spin order in this
phase is shown in Fig.~\ref{fig:figure2}. Indeed, the AFM ordering of the magnetic moments on the Ce sites
has been inferred from low temperature specific heat measurements~\cite{G_Chen:2008}. Furthermore, neutron diffraction
measurements also revealed an AFM order, at low temperatures, of the Re magnetic moments in ReOFeAs
 for Re = Ce, Pr, and Nd.~\cite{Zhao:2008, J. Zhao:2008,Qiu:2008} 

 The electronic density of states (DOS) in ReOFeAs in the phase \{Fe: c-AFM; Re: z-a-AFM\}  
 is shown in Figs.~\ref{fig:figure3} and ~\ref{fig:figure4}, as calculated within GGA and GGA+U, for
Re=Ce, Pr. Note that the GGA 
calculation produces a wide Re 5d-band and a narrow Re 4f band both crossing the Fermi level;
 the DOS at the Fermi energy is 
dominated by 4f states, making ReOFeAs a metal. With onsite Coulomb interaction taken into account, ReOFeAs becomes a Mott 
insulator. By examining the DOS plots, it is noted that within GGA+U, the Re 4f band splits and moves away from 
the Fermi level.
Photoemission experiments should reveal the positions of these bands, providing a check on the validity of the results
obtained in these calculations.

\section{\label{sec:conclusion}Conclusions}
In conclusion, the DFT calculations indicate that the Fe magnetic moments in ReOFeAs, Re = Ce, Pr,  adopt a collinear 
antiferromagnetic order, similar to that in LaOFeAS. Whereas the La ion in LaOFeAS is nonmagnetic, the
Re ion in ReOFeAs carries a magnetic moment due to its localized 4f electrons. Within GGA+U, we show that
the Re magnetic moments also adopt an antiferromagnetic order similar to that adopted by the Fe ions. However,
while the Fe ions in the FeAs layer all lie in one plane, giving rise to a collinear AFM order, with stripe-like
pattern, the Re ions in
the ReO layer lie in two different planes surrounding the oxygen ions plane. In each Re plane, an Re ion is surrounded
by four Re ions that have spins opposite to the spin of the central ion. Viewed in this light, we can say that
in the ground state, within each Re plane, the spin order is simply antiferromagnetic, where each spin-up site is
surrounded by 4 spin-down sites in the same plane. However, for a given spin-up Re ion in a given Re plane, its four NN Re ions
all lie in the other Re plane of the ReO layer; of those four NN Re ions, two will be spin-up and two will be spin-down.
If we connect each spin-up Re site to its NN spin-up sites within a given ReO layer, we will end up with a zigzag
chain running perpendicular to the Fe spin stripes in the Fe-plane. A similar zigzag chain is obtained if we connect 
NN spin-down Re sites within any one ReO layer.


\end{document}